# Probing of local ferroelectricity in BiFeO$_3$ thin films and (BiFeO$_3$)$_m$(SrTiO$_3$)$_m$ superlattices

R. Ranjith, U. Lüders and W. Prellier[1]

*Laboratoire CRISMAT, CNRS UMR 6508, ENSICAEN*

*6 Bd Maréchal Juin, 14050 Caen Cedex, France*

A. Da Costa, I. Dupont and R. Desfeux

*Université d'Artois, Unité de Catalyse et de Chimie du Solide, CNRS-UMR 8181,*

*Faculté des Sciences Jean Perrin, Rue Jean Souvraz, SP 18, 62307 Lens Cedex, France*

Abstract

Ferroelectric BiFeO$_3$ thin films and artificial superlattices of (BiFeO$_3$)$_m$(SrTiO$_3$)$_m$ (m~ 1 to 10 unit cells) were fabricated on (001)-oriented SrTiO$_3$ substrates by pulsed laser ablation. The variation of leakage current and macroscopic polarization with periodicity was studied. Piezo force microscopy studies revealed the presence of large ferroelectric domains in the case of BiFeO$_3$ thin films while a size reduction in ferroelectric domains was observed in the case of superlattice structures. The results show that the modification of ferroelectric domains through superlattice, could provide an additional control on engineering the domain wall mediated functional properties.

[1] prellier@ensicaen.fr

# 1. Introduction

Multiferroic materials, i.e. a material that possesses more than one ordering (Ferroelectric, Ferromagnetic or Ferroelastic) simultaneously, have been the object of fundamental studies and identified as a potential technological applications [1,2]. $BiFeO_3$ is one among the few simple perovskite system, exhibiting multiferroic properties and has been studied extensively in the past few decades [3]. In bulk form $BiFeO_3$ shows a magnetic Neel temperature of ~673K and a ferroelectric Curie temperature of ~1093K. It exhibits a complex cycloidal magnetic ordering of wavelength around 64nm which gives rise to an antiferromagnetic (AFM) behavior [3]. The ferrolectric polarization in $BiFeO_3$ is expected to arise due to the structural distortion induced by the $6s^2$ lone pair electrons of the $Bi^{3+}$ cation [4]. At room temperature, a polarization ranging from ~50μC/cm$^2$ to ~110μC/cm$^2$ has been observed in thin films along the <001> and <111> directions, respectively [5].

In addition to the relatively large polarization, the ferroelectric (FE) domains of the $BiFeO_3$ thin films have been studied extensively [6-8]. Different kinds of FE domain pattern including stripe domains and fractal dimensions have been observed [7,8]. The FE domains of single layer $BiFeO_3$ thin films were found to be larger than the conventional FE systems for a given thickness of an epitaxial thin film [8]. The observed correlation in the domain pattern size and the periodicity of FE domains with the FM domains strongly suggests a multiferroic coupling at the domain walls, which was recently demonstrated on single $BiFeO_3$ layer [8,9]. Hence, the understanding of the domain dynamics and the control over their size and periodicity could facilitate the external tunability of the domain wall mediated coupling.

However, there are certain limitations in utilizing $BiFeO_3$ for device applications [10]. In particular, the large ferroelectric coercive field and high leakage are considered as major limitations [10]. Owing to the high sensitivity of the quality of $BiFeO_3$ thin films on the process parameters, such leakage behavior has been attributed to the presence of additional phases like bismuth oxide and iron oxide [11]. Even without alternate phases, the high DC leakage could be attributed to the presence of oxygen vacancies and the reduction of small fraction of $Fe^{3+}$ cations to $Fe^{2+}$ cations [11,12]. As a consequence, numerous studies have focused on the reduction of this leakage and understanding the leakage mechanisms in $BiFeO_3$ thin films [11,13,14].

Among the various approaches to reduce the leakage current, recently a superlattice with a combination of $BiFeO_3$ and $SrTiO_3$ has been proven to be a useful approach to improve the leakage behavior of $BiFeO_3$ [13,14]. In this work, different series of artificial superlattices structures made of $BiFeO_3$ and $SrTiO_3$ layers was fabricated using the pulsed laser deposition (PLD) technique. The room temperature remnant polarization ($P_r$), leakage current density (J) and their periodicity dependence on the periodicity of the $(BiFeO_3)_m(SrTiO_3)_m$ (m=1 to 10 unit cells) superlattices were studied. The remnant polarization and leakage current density was optimized for a periodicity range of ~ 20-60 Å. The ferroelectric domain structure of both the single layer and superlattice structures was analyzed by piezo force microscopy.

**2. Experimental**

Thin films of $BiFeO_3$, $SrTiO_3$ and their superlattices were grown on (001) oriented $SrTiO_3$ substrates (CrysTec, Germany), at 700°C at an oxygen pressure of 20 mTorr using a multitarget Pulsed Laser Deposition technique. The deposition rates (typically ~ 0.1 Å/pulse) of $BiFeO_3$ and $SrTiO_3$ were calibrated individually for each laser pulse of energy density ~ 1.5 J/cm$^2$. The superlattice structures were synthesized by repeating the bilayer consisting of m unit cells thick $BiFeO_3$ layer and m unit cells thick $SrTiO_3$ layer, with m taking integer values from 1 to 10, keeping a constant total thickness of the superlattice equal to 1200Å. A series of superlattices with periodicity varying in the range of ~8-80Å were fabricated. Prior to the growth of superlattice, a bottom electrode of $LaNiO_3$ (800Å) was deposited at 700°C at an oxygen pressure of 100 mTorr. Gold pads of 400x400μm$^2$ dimensions (physical mask) were sputtered on top of the superlattice structures and on top of LaNiO$_3$ regions unexposed to the superlattice deposition. The fabricated heterostructures were characterized in a metal-insulator-metal configuration to study their ferroelectric polarization and leakage current.

Surface morphology and ferroelectric domain structure of both BiFeO$_3$ thin films and (BiFeO$_3$)$_m$(SrTiO$_3$)$_m$ superlattice structures were studied using a modified commercial Atomic Force Microscope (AFM) (Multimode, Nanoscope IIIa, Digital Instruments) [15]. FE domain imaging was performed using platinum/iridium coated silicon tip with spring constants of ~2.5 N/m. Local piezo phase measurements, out-of-plane (Vertical PFM - VPFM) and in-plane (Lateral PFM - LPFM) measurements were performed to highlight the direction of polarization.

### 3. Results and Discussion

Phase pure BiFeO$_3$ (BFO) and (BiFeO$_3$)$_m$(SrTiO$_3$)$_m$ superlattice structures were grown epitaxially on <001>-SrTiO$_3$ (STO) substrates [13]. Epitaxially grown, phase pure BiFeO$_3$ thin films exhibit an out-of-plane lattice parameter of ~ 4.010Å, which is in close agreement with previous reports. In the case of superlattices, the calculated out-of-plane average lattice spacing revealed that the BiFeO$_3$ in the superlattice structure is under an out-of-plane tensile strain. On increasing the periodicity, the strain relaxes and a shift in the average lattice spacing towards that of the SrTiO$_3$ substrate was observed.

Typical polarization –electric field (P-E) loop of a (BiFeO$_3$)$_5$(SrTiO$_3$)$_5$ superlattice ($\Lambda$ ~ 40 Å) measured at different frequencies is shown in Fig .1(a). The coercive field (E$_c$) of the superlattice structure is ~35kV/cm and is independent of the measured frequency ranging between 1 and 10 kHz. The remnant polarization (P$_r$) obtained from the polarization measurements were in good correlation with the PUND analysis carried out on the same structure.[13] The details of the polarization studies of (BiFeO$_3$)$_m$(SrTiO$_3$)$_m$ superlattice structures with different periodicity is reported elsewhere [13]. The frequency-independent behavior of $P_r$ shows that the observed polarization is intrinsic to the superlattice structure and does not arise from mobile charges or other extrinsic effects, such as leakage current. Since the $P_r$ values observed from both the (P-E) loops and PUND measurements are consistent, we believe that the FE behavior is intrinsic to the superlattice structures.

The leakage mechanism was also analyzed. The leakage current density of the $(BiFeO_3)_m(SrTiO_3)_m$ superlattice structures is dominated by the bulk limited Poole-Frenkel emission in which the emission of charge carriers trapped in the defect centers contributes to the conduction process [16]. Figure 1(b) shows the Poole-Frenkel fitting of the $(BiFeO_3)_m(SrTiO_3)_m$ superlattice structures for m = 5, recorded at different temperatures. The fitting is reasonably good for a wide range of voltage and temperature. The validity of the mechanism could be verified by the magnitudes of the characteristic physical entities derived from the curves. The values of the high frequency dielectric constant and the refractive index derived from the PF type of conduction in the temperature range of 300 – 383K are 5.7 - 7.1 (6.5 for $BiFeO_3$ bulk [17]) and 2.3 - 2.6 (2.5 for bulk $BiFeO_3$ and 2.2-2.6 for bulk $SrTiO_3$ [18]), respectively. The observed values correlate well with the intrinsic material properties of $BiFeO_3$ and $SrTiO_3$ available in the literature and with reported values for the single layer $BiFeO_3$ thin films [10]. In the case of single layer $BiFeO_3$ thin films, the defects are expected to originate from the oxygen vacancy formed due to the mixed oxidation state of $Fe^{2+}$ and $Fe^{3+}$ cations [10]. In the case of superlattice structures, in addition to the oxygen vacancies, the defects could arise from the high trap densities due to the strain fields at the interface like misfit dislocations [19]. Even more, a distribution of shallow traps with low activation energies could be expected at the interface between $BiFeO_3$ and $SrTiO_3$. The details of the analysis of leakage current of the $(BiFeO_3)_m(SrTiO_3)_m$ superlattices in light of various existing models can be found elsewhere [20].

The inset of Figure 1(b) shows the variation of room temperature leakage current density with the superalattice periodicity. In spite of the reduction in the leakage current on increase of periodicity (m≥5), a reduction in the polarization was also observed in the case of $(BiFeO_3)_m(SrTiO_3)_m$ superlattice structures.[13,20] In addition to that, the macroscopic polarization measurements on single layer $BiFeO_3$ films were dominated by the leakage current and hindered the measurements. Figure 2(a) shows the surface morphology of a $BiFeO_3$ single layer thin film; the AFM image was simultaneously recorded with the out-of-plane domain (VPFM) image (figure 2(b)). The image shows the presence of smeared square type grains with a grain size between 100-150 nm with an average roughness of ~ 4nm. The observed morphology and roughness are in consistent with earlier studies reported for films of similar thicknesses.[8] The morphology is in consistent with the existing literature of BFO thin films synthesized under similar process conditions and thicknesses [11]. In the present case, these grains exhibit a clear piezo response demonstrating the absence of conducting impurity phases. The aforementioned observation was further confirmed by the measurement of local piezoloops performed over the surface of the randomly selected grains. [21] In strict sense, the observed domains are piezoelectric domains and the black, white and grey regions represent the force experienced by the probe tip due to upward polarization, downward polarization and intermediate polarization [6-8]. Ideally the dark regions should correspond to the polarization pointing either up or down and the white regions corresponding the polarization either down or up, normal to the film surface respectively in a domain image observed in the VPFM.[6]

Figure 2(c) & (d) shows the out-of-plane domains image and the in-plane domains image respectively on the same sample, but on a larger scan area. Figure 2(e) shows the phase switching (~180º) with applied field, which shows that the observed image contrasts are due to the ferroelectric domain patterns. The combination of the out-of-plane image and the in-plane domain images could give a better understanding regarding the orientations of polarization in a given system. Similar to the out-of-plane image, the dark and white regions in an in-plane image correspond to left and right polarization. The grey regions present in both the images reveal the overlap of the different orientations of polarization in both the vertical and in-plane forces experienced by the tip [7]. This is expected in the case of BiFeO$_3$ thin films whose polarization has been observed to be ~110μ/cm$^2$ along the [111] direction. Hence, the projection of polarization along the [111] direction on the (110) and (100) planes of a given unit cell of BiFeO$_3$ is expected to overlap in both the out-of-plane and in-plane domain image leading to grey regions which has already reported earlier in the case of stripe like domains and other ferrolectric systems [7].

Figures 3(a and b) show the surface morphology (a) of a (BiFeO$_3$)$_{10}$(SrTiO$_3$)$_{10}$ superlattice structure whose top layer is SrTiO$_3$ and out-of-plane domain image (b) of the same area. The surface image clearly reveals spherical grain morphology with a surface roughness of around ~ 1nm. In comparison with the single layer BiFeO$_3$ thin film a reduced domain size and domain pattern is observed. Recent studies reveal the analogy between the magnetic domains and the ferroelectric domains, and as a result the domain wall formation energy is expected to be larger.[8] Hence, also larger ferroelectric domains in the case of BiFeO$_3$ than conventional ferroelectrics is expected [8,9]. The correlation in the ferroelectric and ferromagnetic domains strongly emphasizes the plausible multiferroic coupling in the domain walls [8,9]. Therefore, the strain assisted modification of domain structure and the domain walls present in the superlattice structures, could offer an additional degree of freedom to alter the properties of the system. The observed variation of domain pattern and the domain size in a (BiFeO$_3$)$_m$(SrTiO$_3$)$_m$ superlattice suggests the plausible external control/modification of the domains, in effect, expected to alter the multiferroic coupling. In the case of a (BFO)$_m$(STO)$_m$ superlattice structure the strain distributed over the whole thickness of the film normal to the interfaces and surface is expected to play a crucial role in determining the domain sizes and the macroscopic polarization of the same. A variation in domain size from 100 – 15 nm on varying the periodicity from 20 – 80Å, respectively, was observed.[21] The quantitative studies of strain distribution and its influence over the FE domains are currently under progress.

**4. Summary and Conclusions**

In summary, single phase epitaxial thin films of $BiFeO_3$ thin films and $(BiFeO_3)_m(SrTiO_3)_m$ artificial superlattice structures were fabricated by the pulsed laser deposition technique. The room temperature leakage current density of $BiFeO_3$ thin films was reduced by two orders of magnitude by employing $SrTiO_3$ layers to form a superlattice structure at certain periodicities. The leakage behavior was observed to be dominated by a bulk limited Poole-Frenkel emission. Ferroelectric domains in the range of 120-15 nm were observed in the case of a single layer $BiFeO_3$ thin film. The observed pattern and the size of the domains are in correlation with the earlier observations. The combination of out-of-plane and the in-plane domain images confirmed the overlap of forces on the probe tip from various polarization directions present in the $BiFeO_3$ thin films. In the case of superlattice structures a variation in domain size from 100 – 15 nm on varying the periodicity from 20 – 80Å, respectively, was observed. Finally, we conclude that the superlattice structures of $(BiFeO_3)_m(SrTiO_3)_m$ provide an opportunity to alter the polarization, the leakage current and also facilitates modifying the domain size, pattern and coercive voltages, which are known to directly influence on the multiferroic coupling in the case of multiferroic $BiFeO_3$ thin films and its applications.

**Acknowledgement:**

This work was carried out in the frame of the NoE FAME (FP6-5001159-1), the STREP MaCoMuFi (NMP3-CT-2006-033221), and the STREP CoMePhS (NMP4-CT-2005-517039) supported by the European Community and by the CNRS, France. Partial support from the ANR (NT05-1-45177, NT05-3-41793) and the Région Basse Normandie thought the CPER is also acknowledged. The authors would also like to acknowledge Dr. L. Mechin, Mr. C. Fur, Mr. J. Lecourt, Prof G. Poullain, Dr. W.C.Sheets and Dr. J. Wang

**Figure Captions:**

**Figure 1.** (a) Polarization hysteresis of a $(BiFeO_3)_5(SrTiO_3)_5$ ($\Lambda \sim 40$Å) superlattice structure at different frequencies. (b) Poole – Frenkel plot of the same superlattice structure at different temperatures. The inset shows the room temperature leakage current density at different periodicity.

**Figure 2.** (a) Surface morphology of a $BiFeO_3$ thin film obtained by atomic force microscopy, (b) the out-of-plane piezo force domain image obtained from the same area (c) the out-of-plane piezo force domain image with a larger scan area (5μm x 5μm), (d) the in-plane piezo force domain image of the same region and (e) local phase piezo loop with applied bias for $BiFeO_3$ single layer.

**Figure 3.** (a & b): Surface image of a $(BiFeO_3)_{10}(SrTiO_3)_{10}$ (($\Lambda \sim 80$Å) superlattice structure and the out-of-plane piezo domain image of the same region.

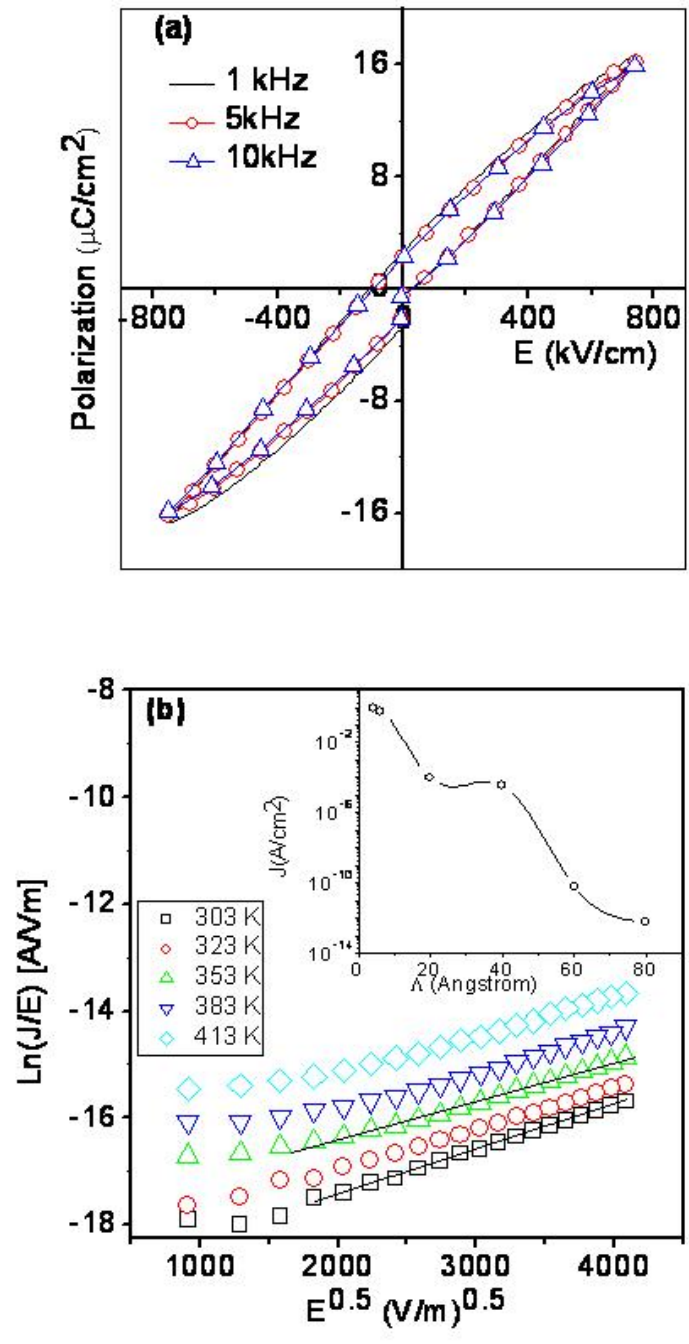

Figure 1

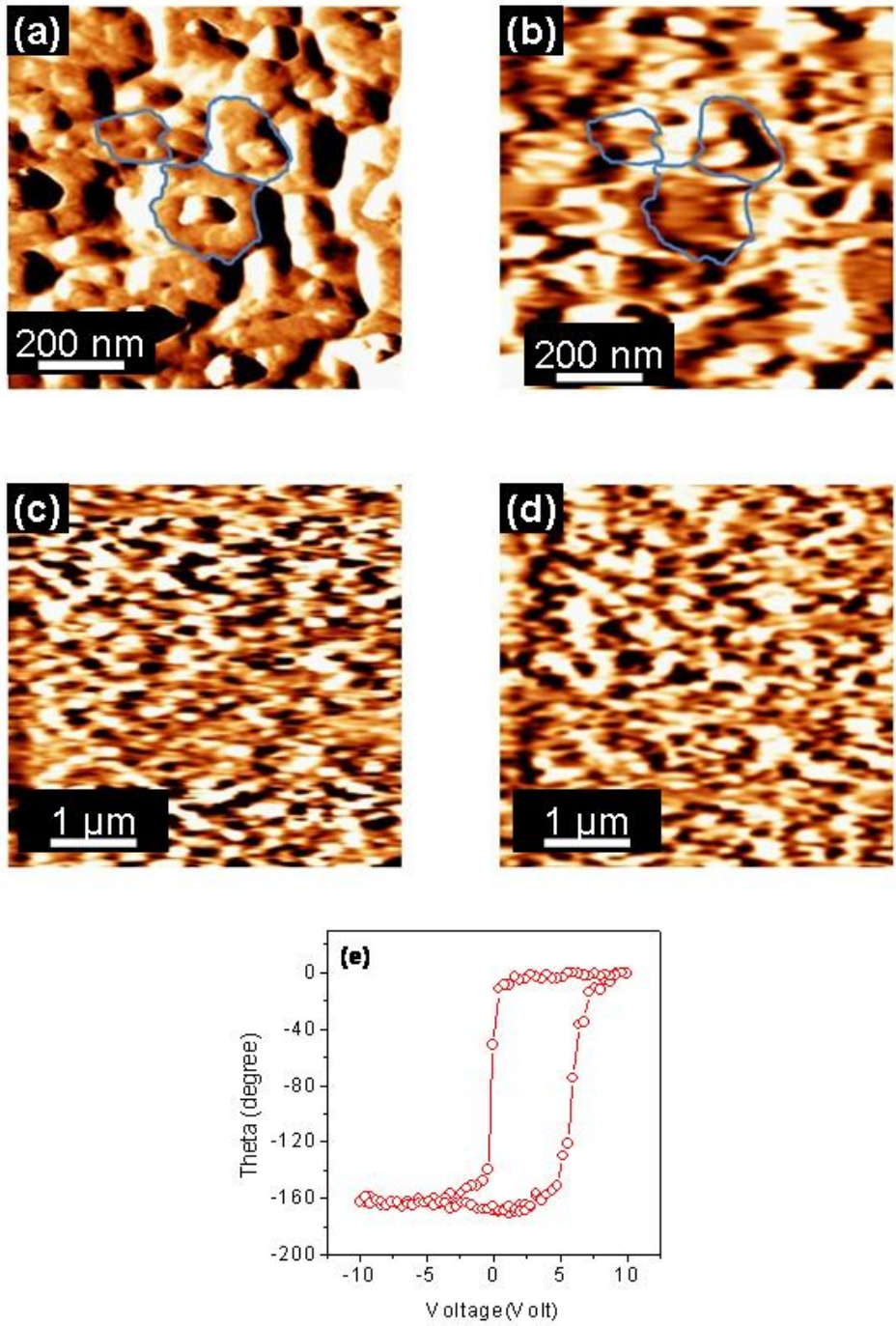

Figure 2

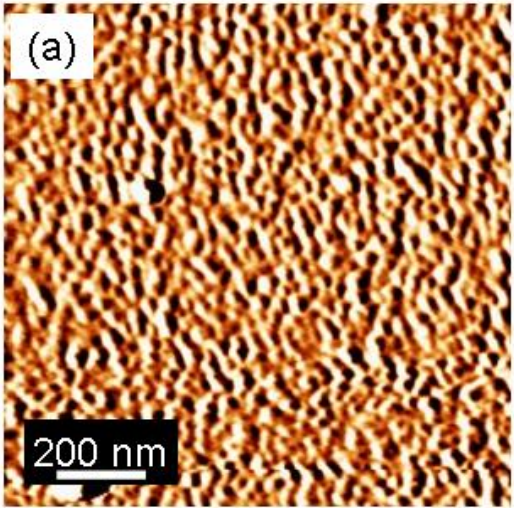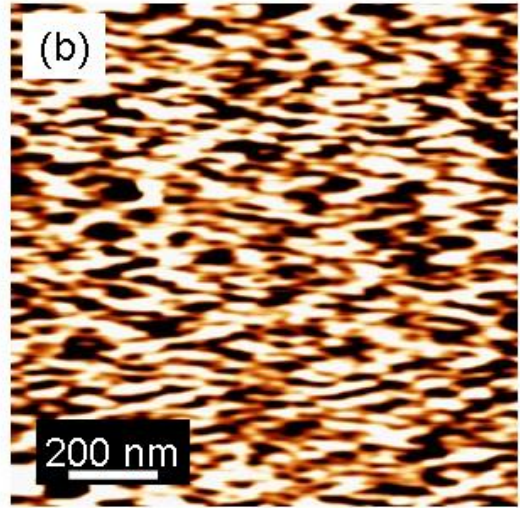

Figure 3

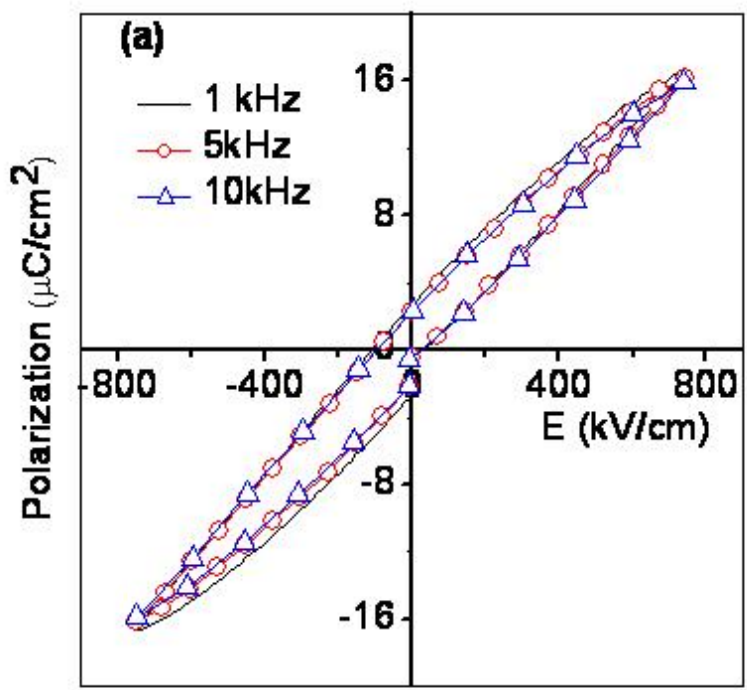

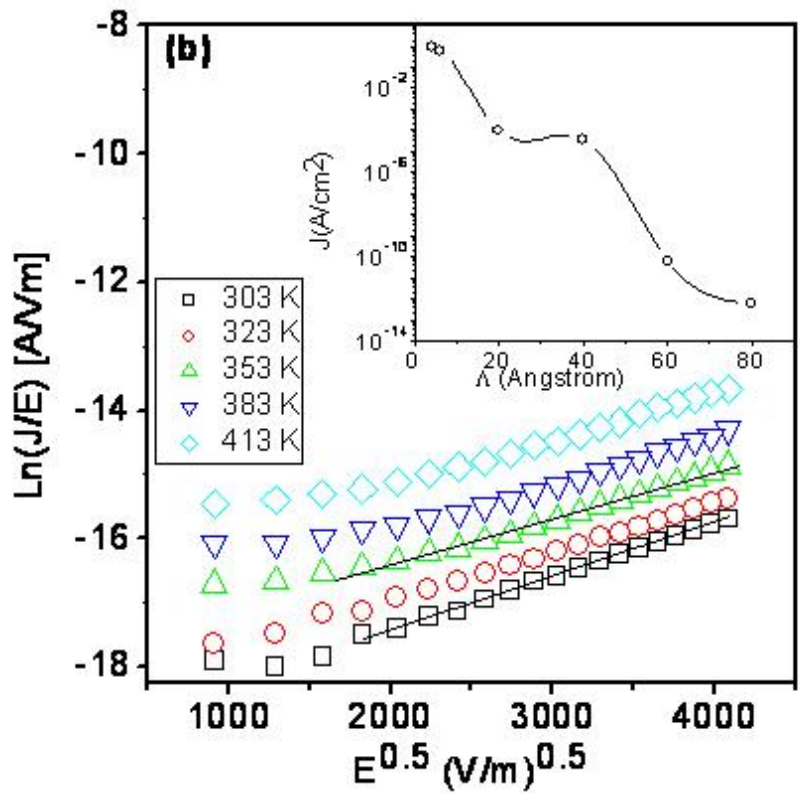

Figure 1

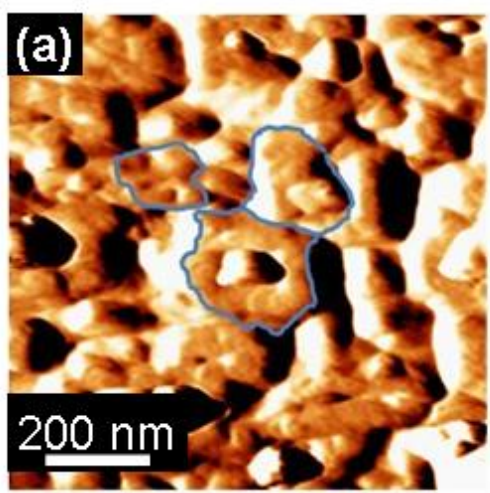
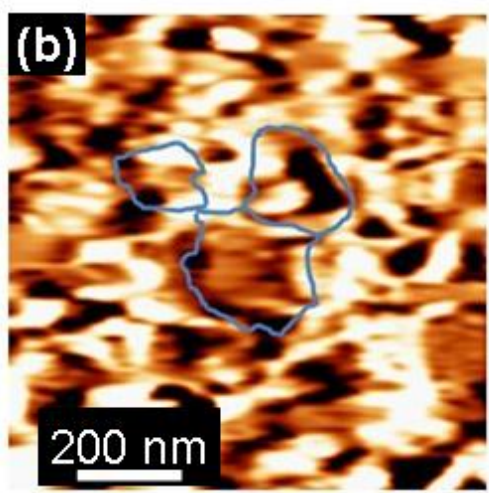
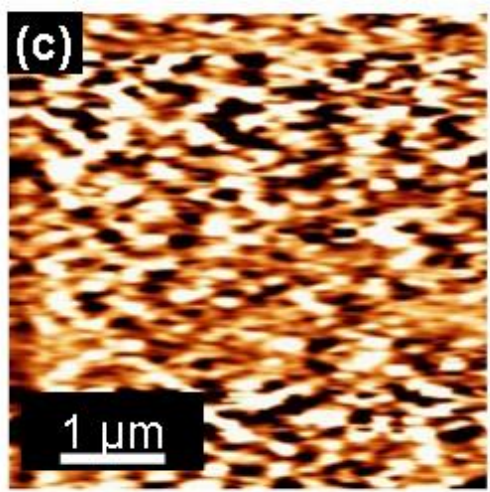
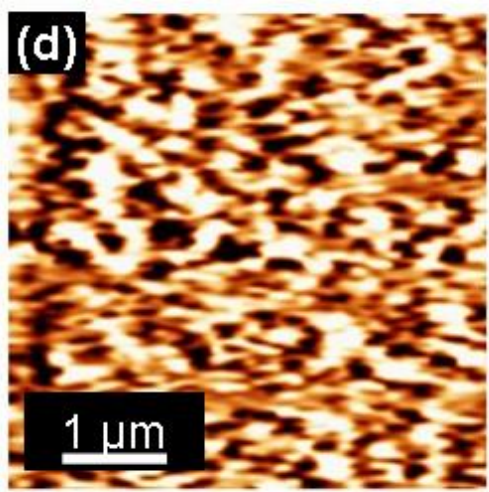
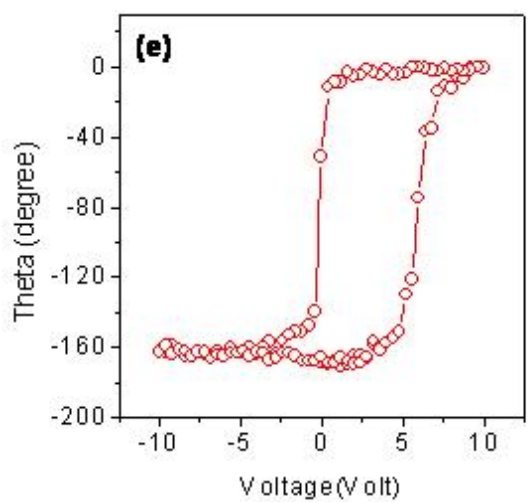

Figure 2

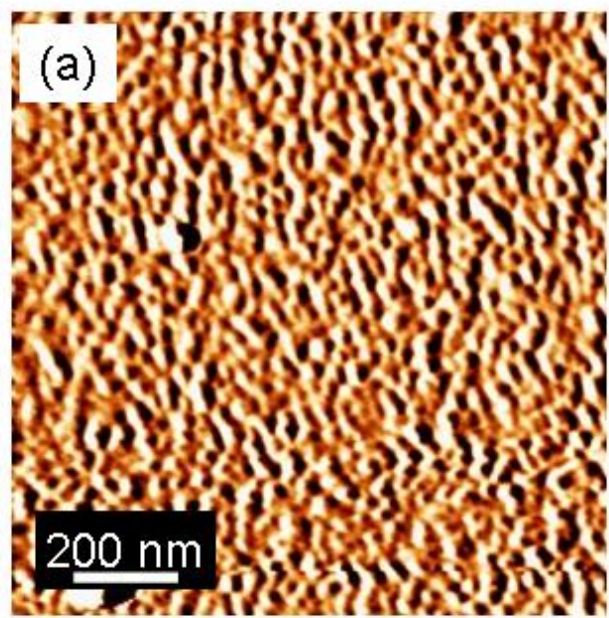 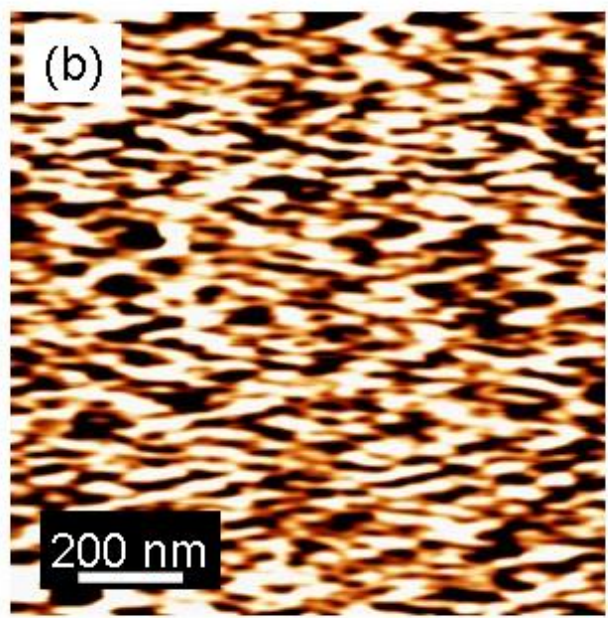

Figure 3